\documentclass[twocolumn,showpacs,preprintnumbers,nofootinbib,amsmath,
amssymb]{revtex4}

\usepackage{graphicx}
\usepackage{amsmath, amsthm, amssymb}

\newcommand{\be}{\begin{equation}}
\newcommand{\ee}{\end{equation}}
\newcommand{\bea}{\begin{eqnarray}}
\newcommand{\eea}{\end{eqnarray}}

\begin{document}
\title{Universality Class of the Reversible-Irreversible Transition
in Sheared Suspensions}
\author{Gautam\ I.\ Menon\footnote{Email: menon@imsc.res.in}} 
\affiliation{The Institute of Mathematical Sciences, 
C.I.T. Campus, Taramani, Chennai 600\ 113, India }
\author{Sriram\ Ramaswamy\footnote{Email: sriram@physics.iisc.ernet.in}
\footnote{Also at: CMTU, JNCASR, Bangalore 560 064, INDIA}}
\affiliation{Centre for Condensed Matter Theory, Department of Physics, Indian Institute of Science, Bangalore 560\ 012, India }
\date{\today}
\pacs{05.65.+b,47.15.G-,47.27.eb,47.51.+j,47.57.-s}

\begin{abstract}

Collections of non-Brownian particles suspended
in a viscous fluid and subjected to oscillatory
shear at very low Reynolds number have recently
been shown to exhibit a remarkable dynamical phase
transition separating reversible from irreversible
behaviour as the strain amplitude or volume fraction
are increased. We present a simple model for this
phenomenon, based on which we argue that this
transition lies in the universality class of the
conserved DP models.
This leads  to predictions for the
scaling behaviour of a large number of experimental
observables. Non-Brownian suspensions under oscillatory
shear may thus constitute the first experimental
realization of an inactive-active phase transition
which is not in the universality class of conventional
directed percolation.

\end{abstract}

\maketitle

The equations of fluid dynamics become time-reversible
in the Stokesian limit where dissipation dominates and
inertia is ignored \cite{stokeseq}.  Neutrally buoyant
non-Brownian particles
in a sheared Stokesian
fluid, do, nonetheless, diffuse \cite{eckstein, leighton,
marchioro, sierou}.  This irreversible behaviour is
the result of an infinite sensitivity \cite{chaos}
to initial conditions.

In ref. \cite{pgbl}
a suspension of neutrally buoyant PMMA spheres (diameter 200 $\mu$m, 
Brownian diffusivity $\sim 10^{-14}$ cm$^2$/s) 
in a Newtonian fluid of viscosity 3 Pa s,
was subjected to
periodic shear $\gamma(t) = \gamma \sin{\omega t}$. 
The Reynolds number in the experiments was less than 
$10^{-3}$, justifying the Stokesian approximation. 
At low volume fraction $\phi$ or $\gamma$ the trajectories
were observed \cite{pgbl} to be reversible, but
at large enough $\phi$ or $\gamma$ they were seen
to be chaotic, irreversible, and consistent with
diffusive behaviour. These two regimes were found to
be separated by a remarkable continuous nonequilibrium
phase transition. 
More recently Cort\'{e} {\it et
al.} showed 
that some initial random motion always
arises upon application of oscillatory shear. 
However,
for small enough $\phi$ and $\gamma$ this motion is
transient, disappearing with a relaxation time $\tau$
that diverges as a power law as the transition is
approached. For large enough $\phi$ and $\gamma$
a nonzero level of activity persists indefinitely.
The universality class of this unique dynamical phase
transition is the subject of this paper.

The physics underlying this transition is well
described by an elegant simulational model 
\cite{corte} of point particles
in a square domain subjected to sinusoidal shear, 
and suffering random displacements
if they come within a specified distance of each other. 
For small $\gamma$
and $\phi$, most particles never meet another, 
and passively follow the
imposed oscillatory strain. Those few particles which
happened initially to be close together will undergo
encounters and move apart, 
and the particle positions will settle down into
one of a potentially infinite number of inactive or
``absorbing'' configurations. In such configurations,
all particles are far enough apart that their relative
positions are not disrupted over a cycle of periodic
strain.  As $\gamma$ or $\phi$ is increased, absorbing
configurations become rarer, so the time to settle
into a quiescent state grows. Across a threshold
strain amplitude $\gamma_c(\phi)$,
the effect
of close encounters propagates throughout the system,
stimulating persistent, global particle diffusion.

The reversible-irreversible behaviour seen in
the experiment is thus explicitly a transition
from a highly degenerate absorbing state to an
active state \cite{corte}.
(An individual
particle is said to be active if it has moved
perceptibly as a consequence of its interaction with
other particles as the suspension is sheared through a
full cycle. )
The directed percolation (DP) transition
\cite{dp1,dp2,hinrichsen} was conjectured to be the
generic transition out of a highly degenerate absorbing state.
Systems like those of \cite{pgbl,corte} in which 
activity is carried by
the motion of a locally conserved quantity such as
particle number, however, are expected to lie in a
universality class distinct from DP \cite{odor,lubeck}.
The key proposal of this paper is that the transition
seen in the experiments should lie in the universality
class of such {\em conserved} DP models, the CDP
class. 

We motivate our model by abstracting what we believe are 
the essential details of
\cite{pgbl,corte}. 
The experiments sample density
configurations stroboscopically, at time intervals
fixed by the periodicity of the imposed strain,
thus focusing only on that component of individual
particle motion which is determined by its interactions
with the other particles.  A particle which does not
have other particles in its close vicinity executes
reversible motion over the time period of the imposed
strain; measured stroboscopically, the particle
is at rest. In regions of larger volume
fraction, 
particle displacements 
in close encounters are in effect
random.

A simple model follows from the
physical picture described above. Consider a lattice
with sites which
can be occupied either by one particle or none (a
vacancy). Particles with no nearest neighbour site
occupied do not move. 
Particles occupying adjacent sites are moved to randomly 
chosen empty sites which
neighbour them. A particle is tagged as active if it
has moved in the previous time step. 
All sites are updated in parallel. 
The fraction of occupied sites, which we shall call
the concentration, is the sole control parameter.
The question of interest is: For generic initial
conditions, is there residual activity in the particles
as $t \rightarrow \infty?$  

The model defined above has been studied earlier  
\cite{jensen,vespignani,rossi}. As the concentration is
increased, the model exhibits an inactive-to-active phase
transition which is {\em not} in the DP universality class
\cite{vespignani,rossi,manna,bonachela}. This is
because the activity, not a conserved quantity,
is carried by particles whose number is locally
conserved. 

\begin{figure}[t]
\begin{center} \includegraphics[width=8.0cm]{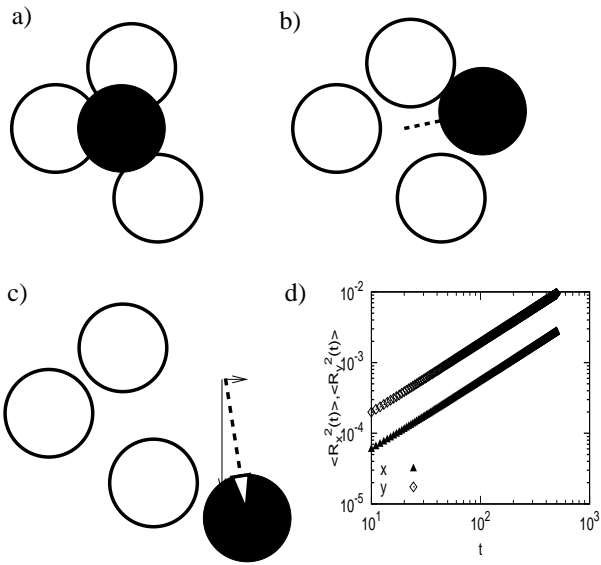}
\end{center}
\caption{
Update of particles in the off-lattice, anisotropic model for
conserved directed percolation.
A particle (black) with one or more other particles in its
interaction radius $R_0$ (Fig. 1(a)) is updated
by first giving it a random displacement in a direction chosen 
so as to correspond to a local decrease of the density
(Fig. 1(b)).  The particle is also further
displaced by a random vector, whose magnitude is chosen
from a uniform zero-mean distribution with different widths in $x$ and
$y$ directions, as shown in Fig. 1(c).  
Fig. 1(d) illustrates the anisotropic diffusive behaviour
of particles above the active-inactive transition
where the mean-square displacement in the $x-$ and $y-$
direction, averaged over all particles, as
$R_x^2 = \frac{1}{N}\sum_i \langle 
\left [R^i_x (t) - R^i_x (0) \right ]\rangle$
The anisotropy $D_y/D_x = 3.8$ in the data shown, comparable to
the values obtained in the experiments.
}
\label{fig1}
\end{figure}

The lattice model can be mapped to a continuum stochastic dynamical 
field theory 
\cite{pastor-satorras,lubeck}: 
\bea
\frac{\partial A}{\partial t} &=& D^A\nabla ^2
A + \mu A - \lambda A^2 + \kappa 
\rho A + \sigma \sqrt{A} \eta \nonumber \\
\frac{\partial \rho}{\partial t} &=& D^\rho
\nabla^2 A
\label{pdes}
\eea
where $A({\bf r},t)$ denotes the activity field and $\rho({\bf r},t)$ the 
local number density of particles. The quantities 
$D^A,D^\rho, \mu,\lambda$ and 
$\kappa$ are constants -- in general depending on concentration --
and $\eta({\bf r},t)$ is a spatiotemporally white 
Gaussian noise. These equations encode the
following: (i) particle motion carries activity to nearby particles; 
(ii) local density promotes local activity through $\kappa$;
(iii) activity gradients produce particle diffusion; and
(iv) noise arises only in regions with activity and, hence, 
with particles.  
Numerical integration of
the equations 
yields exponents which coincide with
the results from the several versions of the lattice
models which have been proposed to exhibit conserved
DP scaling\cite{dornic}. 

The average activity $\langle A \rangle$ is defined in the lattice model
\cite{lubeck} as the time-averaged fraction of active particles, and in the
continuum model \cite{dornic} as the steady-state mean of the field $A$.  It is
found that $\langle A \rangle$ is zero below the transition and non-zero above
it, with a continuous but non-analytic onset: $\langle A\rangle \sim (\rho -
\rho_c)^\beta$ where $\beta$ is a universal critical exponent. 
Close to this transition, fluctuations in the activity are correlated over
distances $\xi \sim (\rho - \rho_c)^{-\nu}$, and times $\tau \sim \xi^z$.
Simulations of the lattice models and the equivalent continuum equations given
above find $\beta \simeq 0.84$ in dimension $d=3$ and $\beta \simeq 0.64$ in
$d=2$ \cite{lubeck}. The correlation length exponent $\nu \simeq 0.59$ ($d=3$)
and $0.79$ ($d=2$) while the dynamical exponent $z \simeq 1.82$ ($d=3$) and
$1.53$ ($d=2$) \cite{lubeck}. The mean-field values for both DP and CDP models
are $\beta = 1, \nu = 1/2, z = 2$. 

How well is the physics of the experiments of \cite{pgbl,corte} incorporated in
the equations above?  It is clear that the detailed trajectories, governed by
the Stokesian dynamics of particles and fluid, are irrelevant to understanding
the transition itself. All that matters is whether the particles return to
their original states upon stroboscopic sampling at the frequency of the
applied strain, or have moved randomly as a consequence of interactions during
the strain cycle. The essence of this physics is adequately included in the
lattice model and its field theoretic translation. It is important that the
physics of the transfer of activity is local -- in the experiments, activity
initiated in a local region of the sample only affects contiguous regions.
Since the experiments are performed in a Couette geometry with a narrow gap $w$
in the gradient direction, the hydrodynamic interaction is highly attenuated on
scales larger than $w$ in the remaining two (axial and circumferential)
directions. We shall comment below on what to expect when all sample dimensions
are comparable. 

Mapping the experiment to the model requires two key assumptions. First is the
reasonable expectation that driving such a highly overdamped system at nonzero
frequency should be irrelevant to the hydrodynamic behaviour once averaged over
the driving period \cite{aghababaie}. Second and perhaps more relevant is our
assumption of isotropic diffusion. The motion induced by close encounters
between particles is not isotropic even in the velocity-vorticity plane. Is
such anisotropy relevant at asymptotically large length- and time-scales? 
This reduces to the question of
whether the anisotropies of the diffusivities for activity and density are
equal at such scales. 

\begin{figure}[t]
\begin{center} \includegraphics[width=8.0cm]{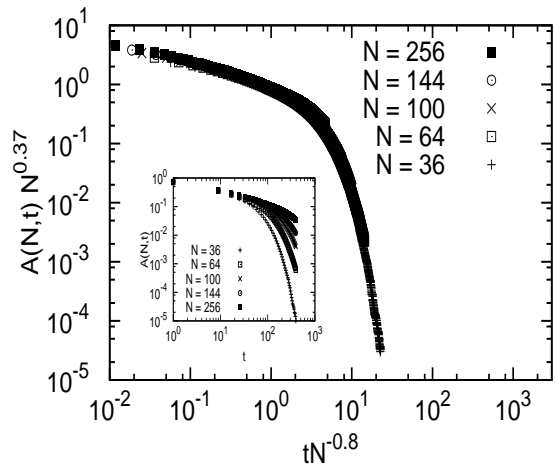}
\end{center}
\caption{Scaling collapse of the fraction of active 
particles $A(N,t)$ for particle numbers
$N = 36, 64, 100, 144$ and $256$, plotted as  
$A(N,t) N^{\alpha z/2}$ {\it vs.} $t N^{-z/2}$.
The data collapse illustrated is obtained for
$\alpha=0.42$ and $z = 1.6$. The inset shows the
bare data corresponding to the scaling plot.
}
\label{fig2}
\end{figure}

We examine these issues through simulations of an off-lattice
model of conserved DP which mimics our stroboscopic
interpretation of the simulation model of Corte et
al. Our model considers $N$ particles in a square
simulation box of area ${\cal A}$, each with an interaction
radius $R_0$, thus defining a  coverage $NR_0^2/{\cal A}$. Every
particle with one or more other particles in its
interaction radius (Fig. 1(a)) is updated in parallel as shown
in Figs. 1(b) and (c), by giving it
a random displacement in a direction chosen so as
to correspond to a local decrease of the density
(Fig. 1(b)). In addition, the particle is also further
displaced by a random vector, whose magnitude is chosen
from independent uniform distributions in $x$ and $y$ with
widths $W_x$ and $W_y$ which are, in general, unequal.
As a consequence, above the threshold, particle motion is
anisotropic and diffusive, as illustrated in Fig. 1(d),
where the mean-square displacement in the $x-$ and $y-$
direction, averaged over all particles is shown separately.

The scaling behaviour of our off-lattice, anisotropic generalization of the CDP
model can be studied through the scaling ansatz, valid at the critical point
\be A(N,t) N^{\alpha z/2} = {\tilde F} (a_N t N^{-z/2}) \label{scaling} \ee
where the factors of $z/2$ follow from the fact that we work at fixed number of
particles $N$, varying $R_0$ so as to maintain the coverage. The inset of Fig.
2 shows the time-dependence of the fraction $A(N,t)$ of active particles as a
function of time at the critical point. This data, obtained by averaging over $10^6-10^7$
independent runs,  is plotted using the scaling
ansatz of Eq.~\ref{scaling} in the main panel of Fig.~2, for an anisotropy
$D_y/D_x \sim 3.5$. Data collapse is obtained for $\alpha = 0.42 \pm 0.02$ and
$z = 1.6 \pm 0.1$, to be compared to the best values obtained from lattice
simulations of $\alpha = 0.423$ and $z = 1.54$. Scaling fits for smaller values
of the anisotropy $D_y/D_x \sim 1.6$, yield exponent values $\alpha = 0.42 \pm
0.01$ and $z = 1.55 \pm 0.05$; for zero anisotropy values within the same range
are obtained. Thus, we conclude that anisotropies comparable to those in the
experiments do not affect the critical exponents of the isotropic CDP transition
and that our anisotropic off-lattice CDP model is in the same universality
class as the lattice model.

The experimental system, in principle, allows us to explore the full range of
dimensionalities and dimensional crossovers in the conserved DP problem,
through the reduction of one or more dimensions of the Couette cell in which
the experiments are done.  If the gap thickness $w$ and radius $R$ of the
cylinder are kept small and fixed and the length $L \gg w, R$, 
the exponents that emerge should be compared to those
of the one-dimensional C-DP model. If instead $R \sim L \gg w$, 
the system is effectively two-dimensional. The experiments
of Cort\'e {\it et al.} are performed in a quasi-two-dimensional geometry in
which the gap between the two cylinders in the Couette cell is the smallest
relevant length scale, corresponding to about 11 particle diameters. 
The growth of relaxation times close to the active-inactive
transition should follow from the scaling exponents outlined above, with 
$\tau \sim  (\rho - \rho_c)^{-1.22}$
for $d = 2$. 
For this exponent, Cort\'{e} \textit{et al.},
find a value close to $1.2$, in encouraging agreement with our
predictions.

Experimentally, Cort\'{e} {\it et al.} find the
order-parameter exponent $\beta$ to be close to the
mean-field prediction of 1, 
but this estimate depends on a
threshold criterion for the presence of
particle motion.
In addition, the experimental 
transition is probably rounded for reasons we discuss 
below. The simulations of the two-dimensional
model proposed by Cort\'{e} {\it et al.} obtain an
exponent of $\beta \simeq 0.45$, probably without 
a detailed finite-size scaling analysis; recent 
related work \cite{mangan} finds $0.59$, closer to 
our predicted value.
The mapping to
CDP provides predictions for several other possible
experiments involving the evolution of activity
under local perturbations in systems  at the critical
point: the average number of active sites $N(t) \sim
t^{-\theta}$ and the survival probability $P(t) \sim
t^{-\delta}$ of the activity after time $t$. Here the
exponent values are $\theta \simeq  0.31$ (d=2) and
$\theta \simeq 0.14$ (d=3) while $\delta \simeq 0.51$
(d=2) and $\delta \simeq 0.76$ (d=3)\cite{lubeck}.
Experiments which involve examining the behavior of the
activity under such local perturbations about inactive
states do appear to be possible\cite{pineprivate}, and
would be an important test of the ideas proposed here.

Some important caveats: 
In a Couette cell with gap $w$, 
the hydrodynamic interaction between particles does not
vanish abruptly at a finite distance but is merely
reduced substantially on scales larger than $w$ in
the remaining directions \cite{diamant,srsed}. 
Small perturbations in particle
positions may then be communicated weakly over
infinite distances. A tiny irreversibility could
thus persist \cite{stone} at the smallest values
of $\gamma$ and $\phi$, and its cumulative effect
over many strain cycles should be observable,
as weak diffusion of the density in the nominally
inactive regime.  Such an additional diffusion
is analogous to a weak
field conjugate to the order parameter in conventional
critical phenomena, turning the sharp transition into
a rapid crossover \cite{diffusion}. 
The experiments of \cite{pgbl,corte} may well be seeing this {\em rounded}
transition from weak to strong irreversibility \cite{jorge}. The
nonequilibrium phase boundary reported in
\cite{pgbl,corte} would then simply be the locus of
points in the $(\gamma,\phi)$ plane below which the
time taken for irreversible behaviour to manifest
itself exceeds the time-scale of the experiment.

The effectively one-dimensional case $L \gg w, R$,
where hydrodynamic screening is expected \cite{diamant}
to be exponential, should show behaviour closest to
a genuine transition.  The long-ranged nature of the
hydrodynamic interaction will manifest itself fully in
the ``three-dimensional'' limit $L \sim R \sim w \to
\infty$; we cannot be sure that the 3-dimensional C-DP
model with local interactions applies to this case. We
note that the Stokesian simulations of PGBL, performed
using periodic boundary conditions and thus in effect
simulating the three-dimensional system, show Lyapunov
exponents which, though small, are still non-zero
even at the lowest shear amplitudes.  Experiments and
simulations which probe the variation of the critical
shear rate for the onset of the transition, with $w$,
$R$ and $L$ varied so as to interpolate between one
and three-dimensional behaviour should illuminate the
role played by the far-field part of the hydrodynamic
interaction and the role of effective dimensionality
in this problem.

In conclusion, we have suggested that the universal
behaviour which should underly the experiments and simulations 
of \cite{pgbl,corte} is to be
identified with the universality class of conserved
directed percolation \cite{lubeck}. This identification leads to
specific predictions for the exponents of the active
to inactive transition in the experiments.  We thus
propose that these experiments constitute 
the first experimental realization of a system in the
universality class of C-DP.

We are grateful to P. Chaikin, L. Cort\'{e} and
D. Pine for many discussions and for freely sharing
the results of unpublished and ongoing work. We thank
D. Pine in addition for asking many probing and useful
questions concerning the applicability of the theory
to the experiments. We thank D. Lacoste, H. Stone,
M. Schindler, P. Ray, J. Kurchan and R. Adhikari for useful discussions. Much
of this work was done when the authors were visiting
the Institut Curie and the ESPCI, Paris under grant
3504-2 of the Indo-French Centre for the Promotion of
Advanced Research.  The DST, India, provided support
through the Centre for Condensed Matter Theory and
Swarnajayanti (GIM) and J C Bose (SR) Fellowships.


\begin{thebibliography}{99}

\bibitem{stokeseq} G.K. Batchelor, {\em An Introduction to 
Fluid Dynamics}, Cambridge University Press, Cambridge (2002)

\bibitem{eckstein} E.C. Eckstein, D.G. Bailey and A.H. Shapiro,
J. Fluid. Mech, {\bf 79} 191 (1977)

\bibitem{leighton} D.Leighton and A. Acrivos, J. Fluid Mech {\bf 177},
109 (1987) 

\bibitem{marchioro} M. Marchioro and A. Acrivos, J. Fluid Mech {\bf 443},
101 (2001) 

\bibitem{sierou} A. Sierou and J.F. Brady, J. Fluid Mech {\bf 506},
285 (2004) 

\bibitem{chaos} E.A Ott, {\em Chaos in Dynamical Systems}, Cambridge 
University Press, Cambridge (2002)

\bibitem{pgbl} D.J. Pine {\it et al.}, Nature {\bf 438},
997 (2005)

\bibitem{corte} L. Cort\'e {\it et al.}, Nature Physics {\bf 4}, 
420 (2008)

\bibitem{dp1} H. K. Janssen, Z. Phys. B {\bf 42}, 151 (1981)

\bibitem{dp2} P. Grassberger, Z. Phys. B {\bf 47}, 365 (1982) 

\bibitem{hinrichsen} H. Hinrichsen, Adv. Phys. {\bf 49}, 1 (2000)

\bibitem{odor} G. Odor, Rev. Mod. Phys, {\bf 76} 663 (2004)

\bibitem{lubeck} S. L\"ubeck, Int. J. Mod. Phys. B {\bf 18}, 
3977 (2004)

\bibitem{janosi} I.M. J\'anosi, T. T\'el, D.E. Wolf and J.A.C. Gallas,
 Phys. Rev E, {\bf 56} 2858 (1997)

\bibitem{jensen} H.J. Jensen, Phys. Rev. Lett {\bf 64}, 3103 (1990)

\bibitem{vespignani} A. Vespignani {\it et al.}, Phys. Rev. Lett 
{\bf 81} 5676 (1998); Phys. Rev. E {\bf 62} 4564 (2000); R. Dickman {\it 
et al.}, Braz. J. Phys. {\bf 30}, 27 (2000)

\bibitem{rossi} M. Rossi, R. Pastor-Satorras and A. Vespignani,
Phys. Rev. Lett {\bf 85}, 1803 (2000)

\bibitem{manna} S.S. Manna, J. Phys. A {\bf 24}, L363 (1991)

\bibitem{bonachela} J.A. Bonachela, H. Chate, I. Dornic and M.A. Munoz,
Phys. Rev. Lett, {\bf 98} 155702 (2007)

\bibitem{pastor-satorras} R. Pastor-Satorras and A. Vespignani,
Phys. Rev. E {\bf 62}, R5875 (2000) 

\bibitem{dornic} I. Dornic, H. Chate and M. A. Munoz, Phys. 
Rev. Lett, {\bf 94} 100601 (2005)

\bibitem{aghababaie} Y. Aghababaie, G. I. Menon and M. Plischke
Phys. Rev. E, {\bf 59}, 2578, (1999)

\bibitem{pineprivate} D.J. Pine, {\em private communication}

\bibitem{stone} We thank Howard Stone and David Pine for a useful
discussion on this point.

\bibitem{diamant} H. Diamant, B. Cui, B. Lin, and S. A. Rice,
J. Phys. Cond. Matt. {\bf 17}, S2787 (2005);
B. Cui, H. Diamant, and B. Lin, Phys. Rev. Lett {\bf 89},
188302 (2002)

\bibitem{srsed} S. Ramaswamy, Adv. Phys. {\bf 50} (2001) 297-341.

\bibitem{diffusion} S. L\"ubeck, Phys. Rev. E {\bf 65}, 046150 (2002)

\bibitem{jorge} A related conclusion is obtained by G. During, D. Bartolo and
J. Kurchan, Phys. Rev. E {\bf 79}, 030101(R) (2009).

\bibitem{mangan} N. Mangan, C. Reichhardt and C.J. Olson Reichhardt, 
Phys. Rev. Lett {\bf 100}, 187002 (2008)
report an irreversible-reversible transition in 
a simulation model for periodically driven vortex systems in two dimensions 
and in the presence of quenched disorder. We believe that our basic
conclusions regarding the universality class of this transition
should apply to these results as well.


\end{thebibliography}
\end{document}